\documentclass[1p,12pt]{elsarticle}

\begin{document}

\begin{frontmatter}

\title{Are We Entering a Paradigm Shift for Dark Matter?} 

\author{James Schombert}
\address{Inst. for Fundamental Science, Univ. of Oregon, Eugene, OR 97403, USA}

\begin{abstract}

While the $\Lambda$CDM framework has been incredibly successful for modern cosmology,
it requires the admission of two mysterious substances as a part of the paradigm,
dark energy and dark matter.  Although this framework adequately explains most of the
large-scale properties of the Universe (i.e., existence and structure of the CMB, the
large-scale structure of galaxies, the abundances of light elements and the
accelerating expansion), it has failed to make significant predictions on smaller
scale features such as the kinematics of galaxies and their formation.  In
particular, the rotation curves of disk galaxies (the original observational
discovery of dark matter) are better represented by non-Newtonian models of gravity
that challenge our understanding of motion in the low acceleration realm (much as
general relativity provided an extension of gravity into the high acceleration realm
e.g., blackholes).  The tension between current cold dark matter scenarios and
proposed new formulations of gravity in the low energy regime suggests an upcoming
paradigm shift in cosmology.  And, if history is a guide, observations will lead the
way.

\end{abstract}

\end{frontmatter}

\section{Introduction}

It has often be argued that science moves forward through a series of leaps or
revolutions (Kuhn 1962).  An existing framework, or paradigm, which had been
successful in the past in explaining a great deal of natural phenomenon, fails to
explain some new observations, or is not providing a path forward in the face of new
observations.  The existing framework needs to be either modified to account for new
observations or simply abandoned (although rarely is the process this simple).  A new
framework would then describe both past and new observations and, at the same time,
lead to new insight and understanding of underlying processes.  Sometimes new
paradigms even lead to new physics.  These leaps forward are often referred to as
paradigm shifts (Cohen 2015).

There have been numerous paradigm shifts with respect to motion (kinematics) in the
history of astronomy.  Based on ideas of the early Greeks, Ptolemy constructed a
mechanical model of the Solar System using perfect circles.  This was replaced by
the Copernicus/Kepler formulation of heliocentric planetary motion, which in turn led
to Newton and the introduction of gravity as the dynamic cause of celestial motion.
Later Einstein, who introduced relativity to deal with motion in the high energy
realm, led to a greater understanding of exotic astronomical systems, such as neutron
stars and blackholes.  The introduction of dark matter as the dominant source of
gravity on galactic scales has extended this paradigm to cosmological realms.  And,
for over thirty years, the cold dark matter (CDM) paradigm has ruled our framework of
cosmology and galaxy formation.  

Here at the beginning of the 21st century, recent discoveries concerning the
kinematics of rotating galaxies has drawn into question many of the basic assumptions
about dark matter.  For example, rotation curves of spiral galaxies strongly indicate
a coupling between baryons and dark matter (Lelli, McGaugh \& Schombert 2016), which
is confusing as dark matter was originally proposed to only interact with matter by
gravity, the weakest of the fundamental forces.  The discovery of these new kinematic
laws (see below) in the low energy, low acceleration region around galaxies may
signal an end to the cold dark matter (CDM) paradigm but promises new physics on the
horizon, if history is any guide.

It is deeply concerning that CDM models and simulations continue to be augmented, in
an ad-hoc fashion, to reproduce new observations to the point of becoming
non-falsifiable as a scientific theory.  This type of behavior in our literature is
one of the signals that we are at a crossroad for a new paradigm that seriously
considers non-Newtonian aspects to motion on galactic scales.  To better illuminate
this issue of an ongoing paradigm shift, we consider the classical, historical story
of one framework overtaking another, the geocentric to heliocentric paradigm shift,
and compare this to the current CDM paradigm.  This tale has the advantage of being
extremely well studied by historians of science as well as being extensively taught
to the younger members of the astronomical community.

\section{The Copernican Paradigm Shift Lesson}

The classic example of a paradigm shift is the transformation in medieval astronomy
from the geocentric worldview to the heliocentric worldview.  This shift, from the
Ptolemaic to Copernican framework, is taught to astronomy students at all levels, and
is drilled into the research community as the baseline example of how science works
to progress forward in terms of understanding how the Universe operates.  They are
many nuances that have been discovered by historians of science concerning the
heliocentric paradigm shift, but what we teach, regardless of complete historical
accuracy, is along the following lines.

The early Greek philosopher's were strongly influenced by the power of mathematics in
understanding Nature, particularly geometry.  The schools of thought surrounding
Plato insisted on developing a cosmological model focused on perfect circles of
motion for the heavens.  This model connected the four primary elements (Air, Fire,
Earth and Water) surrounding the Earth with the seven planetary sphere's resulted in
a framework where there was linear motion below the Moon's sphere (resulting in ``natural"
motion being motion that returns to one's sphere) plus perfect circular motion in the
heaven's (trans-Lunar).

This framework had the advantage of appealing to common sense (the sky sure looks
like a half-sphere) while also providing a pure geometric view of the heavens, a
clean spherical mechanism.  The details of actual planetary motion are ignored, and
the focus was on perfect shapes around an egocentric center (i.e., us).  The later
combination of Plato/Aristotle philosophies with Christian doctrine formed a complete
cosmology/theology of planetary motion and human morality (so-called scholasticism,
i.e., see Dante's cosmology).

Of course, the devil is in the details (pun fully intended) and attempts to apply
this geocentric framework quickly ran into difficulties with respect to planetary
retrograde motion.  As we teach our students, retrograde motion drives Ptolemy (c.
150 AD) to develop our first astronomical ``kludge", the ad-hoc addition of epicycles
on deferents in order to ``save the phenomenon" of perfect circles.  The Ptolemy
framework is horribly convoluted, (thus, violating Occum's Razor), but became highly
accurate (due to its computational flexibility).  Lack of observed parallax, plus a
poor understanding of inertia, meant the Ptolemy framework remained unchallenged for
many, many years (about 1,500).

The Copernican framework is developed at the beginning of the Renaissance for
primarily cultural reasons as there was nothing inherently inaccurate about Ptolemy's
model and there were no new observations that would have caused the Ptolemaic
framework to be falsified (using modern terms).  There was a general uneasiness
concerning the orbital behavior of the inferior planets, (Mercury and Venus always
being near the Sun, regardless of their orbital center being around the Earth),
however there was no computational flaw to the geocentric model.  While the
Copernicus heliocentric framework is considered to be the standard example of a
paradigm shift, it actually contained nothing superior as a computational tool since
it also continued to use circles, rather than ellipses, and required similar epicycle
kludges to correctly predict the positions of planets in the sky (in fact, the
Ptolemy model continued to be used for centuries to construct almanacs).  The heliocentric
framework was accepted as a ``mathematical fiction" in many scholarly circles at the
time.  Thus, we have two frameworks, both making predictions, many of these
predictions (such as parallax) were outside the technology of that time.

As we teach, along comes Galileo with his new technology, the telescope.  Simple
observations falsify (in Popper's words) the geocentric framework (but do not ``prove"
the heliocentric framework, which is also incorrect with its use of perfect circles).
Kepler's use of ellipses completes the kinematic description of Copernicus (removing
the ad-hoc computational complexity of Ptolemy's framework).  Newton completes the
dynamic framework with gravity, and his laws of motion, plus demonstrating that
Kepler's ellipses are required for an inverse squared force law (and giving us
calculus as a new computational tool).

Again, there were many nuances to this story, but this is the way we teach it to our
students:  

\begin{enumerate}[Step 1.]
\item Two competing paradigms (geocentric vs heliocentric),
\item dramatic observations (Galileo),
\item one framework is falsified,
\item a significant leap forward in our understanding of the Universe
\end{enumerate}

This model of how science works is referenced in ways too numerous to count, even in
circumstances where the exact details may not apply.  The highly visible nature of
this example is also due to the large number of changes in astronomy, physics,
biology that were occurring at roughly the same time in Western Europe along with so
many new technologies (e.g., the microscope and telescope).  The result of many of these
new ideas was a decoupling with the concepts of Aristotle and other Greek thinkers,
opening a path to novel interpretations of Nature.  No one doubts this was a defining
moment in the history of astronomy and the knowledge gained in later years can be
traced back to this paradigm shift.

\section{The Dark Matter Paradigm}

The introduction of dark matter as the primary driver in galaxy kinematics has been well
documented in many review articles (see Sanders 2010).   Early work can be summarized
as the discovery of excess motion (either cluster velocity dispersions or flat
rotation curves) without corresponding luminous matter.  Thus, the so-called ``missing
mass" problem is more correctly phrased as the ``missing light" problem (Oemler 1988)
and initial searches were focused on identifying the missing light as very faint, or
even dark, matter as an obvious solution to the kinematic observations.  And, in many
ways, the assumption of dark matter ``saves the phenomenon" of Newton+Einstein
framework much like epicycles saved Plato's framework.

However, this is not an accurate comparison.  For epicycles, while perhaps
mathematically necessary, do not carry the same weight as the scientific hypothesis
of non-luminous matter to explain the deduced gravitational influence on and within
galaxies.  In addition, there have been many instances of ``missing matter" being
found through gravitational influence (the most famous, of course, is the discovery
of Neptune).  Epicycles, on the other hand, were a convenient mathematical tool with
no history in astronomical calculations.

Despite this difference, the dark matter hypothesis has been particularly difficult
for astronomers to work with as it is central to our frameworks that astronomical
objects describe a reality distinct from the appearance of things.  This is a form of
instrumentalism, a view where frameworks are not necessarily true (in a mathematical
sense) but are accurate and useful fictions to perform computations and make
predictions about astronomical observations.  This works well on phenomenon ranging
from interstellar gas clouds made of atoms and molecules to galaxies made of
point-like stars.  However, dark matter, by definition being non-luminous, is not
sampled directly by our telescopes.  And scenarios using existing, but mostly
non-luminous, astronomical objects (neutron stars, blackholes, black dwarfs)
immediately ran into conflicts with other, well-established astronomical parameters
(such as the age of the Universe).  In other words, dark matter was a useful fiction
with respect to known astronomy, but it's framework was outside known astronomical
objects.  And, unlike the debates of the reality of objects in the microscopic world
(e.g., quantum fields), objects in cosmology are, by definition, macroscopic and must
entail the same realism as known astronomical objects (i.e., stars).  There were only
a very limited number of speculative astronomical objects that satisfied the
characteristics of dark matter (i.e., invisible) and did not violate existing, well
established paradigms (such as the theory of stellar evolution).

The immediate solutions to a non-astronomical dark matter component is either 1) a
known particle with surprisingly high mass or 2) an unknown particle with
an unsurprisingly high mass.  Attention immediately fell on the neutrino, with an
early indeterminate mass.  However, the family of neutrinos were soon shown to have
insufficient mass to account for dark matter.  Thus, research attention was directed
to a new particle, assumed have a very high mass and a uniform distribution needed to
account for motion in the halos of galaxies (cold dark matter, CDM, cold in order to
gravitationally collapse early to form large scale structure).  The loss of the
Superconducting Super Collider in the mid-90's released a large number of resources
(i.e., people) resulting in a great deal of theoretical speculation about a new dark
matter particle.  However, decades of searching and speculation have produced no
tangible object or working framework for a dark particle.  While we are amazed at the
ingenuity of our theoretical community, it is almost impossible to read recent work
in this area without experiencing a sense of desperation.

As critical as astronomers have been to the dark particle enterprise (perhaps a little
jealousy that dark matter had shifted from its astronomical focus and funding), there
are many examples in the history of creative thinking where scientists certainly did
not derive their theories from data.  And one could argue that the discovery of
``missing mass" is the kind of experimental push to framework building much like the
photoelectric effect was to quantum physics.

Meanwhile, after it became increasingly obvious that dark matter was not going to be
in the form of some astronomical object with an origin in baryonic material (brown
dwarfs, neutron stars, blackholes, etc.), the main focus of astronomical
investigations became to gathering more data in order to better outline the dark
matter phenomenon.  There was less an interest in explaining what dark matter was,
and more an emphasis on how dark matter contributed to basic astronomical processes,
such as galaxy formation.  More data about events is the core of the philosophy of
empiricism, but empiricism by itself lacks predictive power.  It did not seem
fruitful to gather more observations of dark matter in the hope of building general
correlations towards a new law of Nature.  Attention was directed to defining the
effects of dark matter on baryonic matter and its influence on cosmological models.

So, currently, one of the greatest challenges with the dark matter hypothesis is how
do we formulate characteristics of things that are not observable?  Being
unobservable is not completely true, as we supposedly see the effect of dark matter
on other matter, and some crude information can be extracted from those observations
(e.g., lensing maps of galaxy clusters).  And it is not uncommon in astronomy to
investigate phenomenon through indirect observations (e.g., blackhole physics from
accretion disk observations).  But dark matter takes us into a very different realm
of indirect observations, especially in that we have no information on the nature of
dark matter itself to guide our investigations.

\section{How does Dark Matter fit into Scientific Realism, is it Predictive?}

Scientific realism is the view that we should believe in the unobservable objects
postulated by our best theories.  But, it also has a negative connotation in that it
implies that common sense reality is an illusion, and that there are layers of
reality that are remote from everyday experience.  This view is not too difficult to
embrace since, even in the laboratory, objects of scientific scrutiny have primary
and secondary properties.  Primary properties are things like mass and size, which
are only altered by changing something fundamental about the object. Those properties
that things only {\it appear} to have are secondary, such as color or taste.  Secondary
properties also have meaning in reality, but only through context (color is reflected
light).  Primary properties are what most scientists consider to be fundamental and
mind-independent.

What we know about dark matter fits poorly into our current view of scientific
realism.  While motion is a primary property of matter, the cause of motion has a
range of possibilities.   Our failed dark matter searches have not even limited the
characteristics of a hypothetical new particle as speculation moves the expected
properties into new realms and often borders on untestable (a classical ``moving the
goalposts" fallacy).  There does not appear to be any particular experiment which can
falsify the dark particle hypothesis as the avenues for speculation are endless.  In
scientific realism, objects exist independent of our minds and senses.  Dark matter
searches seem to presume objects that are not only independent of our senses, but are
also outside our ability to sense them.

Ernst Mach, a late 19th century physicist who strongly influenced Einstein, argued
that science should only concern itself with what was observable, and the function of
natural laws is to systematize the correlations of our observations.  Mach's hope was
to develop a functional system based on fundamental concepts (i.e., independent of
observations) combined with practical correlations (determined in a deductive or
inductive manner, so-called foundationalism).  Dark matter searches seem to reverse
this philosophy as there is very little in the Standard Model to suggest a dark
particle.  These searches also hope for new physics that depends on the discovery of
such a particle to spur extension of the Standard Model into a new framework much
like the many historical examples of new discoveries leading the way to Maxwell's
formulation of electromagnetism and Einstein's development of relativity.  Again,
this seems very un-Mach-like.  While one can never have absolute positive grounds to
believe theoretical entities, like dark matter, no matter how empirically successful
their theories, it still begs the question of whether they are real.

The state of dark matter research closely resembles the state of physics in the early
19th century concerning the caloric model versus the kinetic theory of heat transfer.
The caloric model was extremely successful in many of its predictions, just as dark
matter is a successful explainer of cosmological observations.  The caloric model was
also powerful in its use with developing new technology (e.g., steam engines).
Mathematically elegant plus being precise guaranteed the popularity of the caloric
model.  And similar to dark matter framework, caloric theory required an invisible
and immaterial substance at its core.  Many of the same justifications were used to
support the caloric research paradigm, in the sense that it was a great ``explainer"
of phenomenon and was ``simple" in its formulation (invoking the short form of
Occum's Razor).  It would evidentially fall in its inability to predict heat from
friction in mechanical devices, such as drills, paving the way for a framework of atoms
and motion, but for decades was the center of the new field of thermodynamics.

The dark matter paradigm has several positive characteristics.  For example, it is
very attractive for its promise of future new physics.  The Standard Model seems
incomplete to some, dark matter is a good excuse to test its boundaries.  For
cosmologists, dark matter is critical to explain the details of the CMB and features
in the large scale structure of the Universe as well as key elements to galaxy
formation.  Dark matter is so ingrained into the current cosmology framework that is
is simple inconceivable that it does {\it not} exist for it is the ``simplest
explainer" of the observations.  However, dark matter lacks any predictive power as a
framework since its basic characteristics are unknown and its use in cosmology is
mostly in what its does {\it not} do (i.e., interact with baryons except by gravity).

\section{Is Dark Matter Testable?}

The obvious answer to the question of whether dark matter can be tested as a
scientific hypothesis is ``of course", once a dark matter particle is detected.  Until
that moment, it is a serious concern that dark matter, as it is framed for scenarios
of galaxy formation, is untestable.  This is due to the fact that with each failed
experiment to identify a dark particle, or define a cross section, the ``goalposts"
for its characteristics are moved.  The theoretical community is strongly committed
to a new dark physics framework and, thus, is open to many avenues of formulation
particularly when there are almost no constraints on the actual characteristics of
dark matter.

The main question then becomes how do we reconcile our theoretical frameworks,
particularly computer simulations, with our observations? Logical positivists
struggled with how to separate the empirical content of theories, the synthetic part,
from the theoretical, the analytic part.  Certainly there is a history in astronomy
that our theories and frameworks guide us in deciding what to observe.  Many an
observational program is designed to explore the predictions from a newly proposal
theoretical concept.  And key to that process was the well-known principle of
falsification (Popper 1959) which was originally developed to define a demarcation
between science and non-science, but has over time evolved as a basic principle
underlying scientific exploration (even though philosophers still struggle with
Popper's ideas, see Gardner 2001).

In Popper's scheme, there are many different kinds of predictions from frameworks,
the most powerful ones being novel predictions (those which are predictions of new
types of phenomenon).  If these phenomenon are previously unobserved, then these
frameworks contain bold conjectures and are especially attractive for observational
investigation.  How then does dark matter fit into our astronomical programs?

The greatest tension between astronomical observations and dark matter is our
understanding of galaxy formation.  Increasing telescope size and wavelength coverage
(i.e., new ground-based systems plus far-IR space telescopes) has made for deeper
studies of galaxies to higher redshift.  This has meant more information on the
epochs associated with the first burst of star formation and the first epoch of
galaxy evolution by mergers.  This allows for detailed comparison between
observations and our theoretical foundation on the physics behind galaxy formation
plus predictions from computer simulations.

While this article is not a review, nor criticism, of how we test and refine our models
of galaxy formation and evolution using computer simulations, it is instructive to
examine how that process is perceived by the community.  There is general agreement
among observers that a good overview knowledge of the predictions from computer
simulations makes for a healthy discussion section in a research article.  This is
perceived as a correctly processed scientific procedure that feeds back into our
theoretical framework by isolating the processes that are important to the various
galaxy formation scenarios.

The advent of advanced computer simulations, combined with comparison to high
redshift observations (i.e., temporal or evolutionary information), has given us the
ability to investigate extremely complex phenomenon and extract consistent
predictions to be tested by observations.  This has forced a change in our view of
knowledge from an old system where frameworks must be certain, subject to proof and
free of error (e.g., Newtons laws of motion) to a more modern view where we apply a
critical and introspective analysis of our frameworks for their value in matching
observed phenomena.  Note that the interaction between computer simulations and
observations is critical, the simulations are constrained by reality, and the
observations can only be interpreted by reference to simulations, which model the
underlying physics.

It should also be noted that the construction of computational frameworks is not a
mechanical process, it is a process laced with creativity and rigorous testing.  By
their nature, computational models are unlike general frameworks.  For example,
frameworks can be falsified by just one negative observation, but most computational
models are not of this type.  Some unfalsifiable principles, like conservation of
energy, are considered part of our knowledge and fundamental physics of a simulation.
But, violation of fundamental physics in a simulation would reveal something else is
wrong, not the principles themselves.  When encountering new physics, the typical
scientist will think up modifications to a theory, or extra assumptions, in order to
save it.

Popper would consider computational models acceptable if they make further
predictions for the framework (i.e., ad-hoc vs non-ad-hoc modifications).  For
example, the orbit of Uranus did not falsify Newton framework for it was assumed one
of the input parameters was wrong (i.e., the influence of Neptune).  If a
computational model has lots of evidence in its favor, and it works, it would be
crazy to abandon it without something better to replace it.  This is how, in some
sense, computational models gain an intellectual inertia as they provide some
understanding to the underlying processes.  However, theories can infect data to such
an extent that there is no gathering of observations that can ever be theory-neutral
and objective.  Strong computational models differ from theory in that there is a
disciplinary matrix (i.e., the education of a scientist) which includes skills that
enable scientist to work within the paradigm.  This restricts the avenues that can be
used to modify computational models (individuals have limited tools in their mental
toolbox).

Scientists adopt all manner of strategies to save theories from refutation.
Especially if a framework 1) has had great deal of success, 2) has dealt with
anomalies in the past, 3) has involved a massive investment in time and education and
4) has consumed extensive resources.  In addition, a scientist who pauses to
examine every anomaly seldom gets significant work done (Kuhn 1962).  However, some
anomalies will not go away and may produce conceptual paradoxes (Ladyman 2002).
Kuhn-style revolutions involve change in the context which science questions are
resolved and often evidence will never be enough to compel scientists to switch
paradigms.  Scientists are often thoroughly committed to their paradigm they are
working within and refuting evidence is unlikely to locate the problem within the
central assumptions that define the paradigm.  The idea that a theory is supported by
community because it is believed is called epistemic relativism and the dark matter
paradigm seems deep in this well of logic. 

Part of the problem for our computational models is that, in attempting to describe
an objective reality, words mean different things in different paradigms.  And, while
scientific knowledge is fallible, partial and approximate, it is still the best
method for making predictions.  Computational models represent our most complex
attempt at achieving this goal.

\section{The Route Forward: Acceleration is the Key Parameter}

History has a clear answer to the dark matter problem, more data.  A particle
detection would be a tremendous leap forward but, despite the particle community's
optimism, the odds are looking worst with each passing experiment.  Dark matter is
still an astronomy problem and astronomical observations have provided the only
useful information on dark matter for the last 50 years.

Information on dark matter has historically been kinematic (although lensing maps
have provided the first look at the distribution of dark matter in clusters).  The
most primary kinematic correlation concerning dark matter is the baryonic
Tully-Fisher relation (bTF, McGaugh et al. 2000).  The bTF demonstrates a correlation
between the total baryonic mass of a galaxy and its total dynamical mass
(presumingly, dark matter plus baryonic matter).  While some correlation is expected
since baryons and dark matter arise from the same density fluctuations in the early
Universe, the scatter in the bTF is completely observational, meaning that if one
knows the total baryon mass of galaxy, you can deduce it's total dark matter mass to
within observational error.  Given all the various physical processes involved in
galaxy formation, that act differently on baryons compared to dark matter (e.g., gas
cloud physics), this relationship is startling.  In addition, the slope of the bTF
contradicts most galaxy formation scenarios using dark matter.

An obvious extension to the bTF is to consider all the information contained spatial
in the rotation curve, rather than a total mass value in a Tully-Fisher fashion.
This is the recently discovered radial acceleration relation (RAR, McGaugh, Lelli, \&
Schombert 2016).  This kinematic diagram relates the acceleration at a particular
radius from the baryonic matter with the total acceleration deduced from the rotation
curve (again, presumingly dark plus baryonic matter).  And, again surprisingly, the
baryonic acceleration is very tightly correlated with the total acceleration, the
scatter being completely within the observational errors.  The meaning of this, new,
kinematic law for rotating galaxies is that if one is given the baryonic value at a
particular radius, you can then deduce the total mass value (dark plus baryonic) at
that same radius.  This is impossible given the dissipative processes needed to form
a rotating disk from a dark matter halo.

While sounding esoteric, these relations are remarkably tight by astronomical
standards and relate the baryon mass to the presumed dark matter mass (both on local
and total scales).  For example, the radial acceleration relation demonstrates that
if one knows the baryonic mass at a particular point in a galaxy you immediately also
know the dark matter mass, in other words the two substances are strongly coupled.
In addition, these correlations indicate the existence of an acceleration scale in a
conceptually distinct manner and they return the same value, to within the errors, of
about 10$^{-10}$ m s$^{-2}$. This ubiquitous acceleration scale is a critical clue
about the missing mass problem.  

The conflict between these relationships and the dark matter hypothesis is that dark
matter is presumed to only interact by gravity (the weakest force) with baryonic
matter. Although there exists weakly interacting dark matter models, the coupling
implied by observations exceeds any dark matter scenario, particular in the low
density regime where gravity is also at its weakest.  In other words, it is a
baseline characteristic of dark matter that it {\it not} interact strongly with
baryonic matter (this is required for the success of dark matter in cosmological
models), and the indicated strong coupling is difficult to explain under any dark
particle hypothesis.  Perhaps more importantly, no dark matter framework predicted
these kinematic relationships (although many have recovered this property after
discovery with modifications to their frameworks, Kuhn and Popper would shame them).

The history of galaxy formation theory is laced with models to understand the
structure of galaxies, and the evolution of structure.  These models use parameters
such as characteristic scale length, density and total mass to compare predictions to
observations (then iterate on the input parameters).  Thus, scaling relations, such
as the ones discussed above, are critical to understanding galaxies.  However, it is
becoming increasing obvious that we are asking the wrong questions in attempting to
understand the origin of these scaling relations.  We tend to ask questions related
to the effects of gravity (for example, scale length asks about the size of a
galaxy under gravity as a function of mass or density).  Instead, our new scaling
relations indicate that acceleration is the controlling parameter.  All the
correlations are linked by a common acceleration scale, not a mass or density scale.
Within the uncertainties of the data, all of them express the same acceleration scale
which is not predicted in our current framework of galaxy dynamics.  If this
acceleration scale is unique and universal, then it represents a major change in our
understanding of how galaxies work and explains the increasing popularity of modified
gravity theories as the problem appears to be focused on how things move in galaxies
rather than missing matter.

\section{Is This a Paradigm Shift?}

The evidence for a paradigm shift with respect to the dark matter framework is, as is
expected from the history of science, driven by new observational correlations
between baryons and dark matter.  If the RAR is as important as we think, than we must
reconsider our dark matter scenarios within these new observations and inspect our
initial assumptions about dark matter as non-luminous and non-interacting (except by
gravity).  We appear to be forced to abandon this pathway as the strong coupling with
baryons is exactly the opposite characteristic as originally proposed for dark
matter.  Thus, CDM paradigm appears to be underdetermined as a working
framework.  Underdetermination means that more than one theory is compatible with the
evidence.  Observations underdetermine a framework when they are insufficient to
suspend judgement on theories.  Often explanations require mere empirical adequacy
(which is defined as simplicity, coherence, elegance), but also must give us more
than just a description (like Ptolemy's model) they need to tell us why in a causal
way.

In this context, our computational models that use dark matter appear to be severely
underdetermined.  They simply have too many variables (i.e., feedback) that are
proposed as ``reasonable" corrections.  This kind of modifying of the basic physics in
simulations leads to a scenario of no or little progress and is reminiscent of what
observers experienced during the ``Hubble wars" in the 80's.  Two competing values for
$H_o$'s were sustained for many years mostly by poor experimental design and a weak
computational framework for understanding galaxies.  Numerous ``reasonable"
corrections to distance scale data made the supporting of widely different values
relatively easy.  The feedback mechanisms in computational models has many of the
same features as epicycles in Ptolemy's framework; reasonable, justified by
observations yet probably completely wrong.

Science is often defined as the process of rational inquiry produced by a cumulative
growth of empirical knowledge.  Growth is measured by the strength of our frameworks.
Thus, a paradigm shift is characterized by 1) emphasis on novel observations, 2)
abandonment of qualitative descriptions for quantitative ones, or 3) the appearance
of novel interpretations.  Adopting a new kinematic framework for rotating galaxies
to explain the bTF and RAR, such as MOND (MOdified Newtonian Dynamics, Milgrom 1983), is
very much like the adoption of the Lorentz contraction to explain the
Michelson-Morley experiment.  The original Lorentz contraction was presented as an
ad-hoc fix to the Michelson-Morley results.  It only became a component to special
relativity long after relativity was first proposed.  MOND may not be correct or the
finished product, but if history is any guide, our new path forward is something that
is MOND-like.  Much like a large jigsaw puzzle, MOND appears to be an edge piece to a larger
framework.  One thing that is clear is that our path forward lies in the low density,
low acceleration realm of galaxies.

Our current frameworks will not exist in 100 years.  While there are individuals who
believe we closing on final theories that will explain everything, they have not been
paying attention for the last few thousand years (definitely not in a Bayesian
fashion).  Certainly a framework that attempts to explain things is desired, but it
is not a goal.  Predictive power is the true goal of science and frameworks that lead
us down new paths are what we really desire.  Frameworks that fail to lead us in new
directions need to be abandoned.  Not because they are wrong in any real sense, but they
simply have served their role in getting us ready for the next step.  The CDM
framework has reached this point.






\medbreak

\noindent \textbf{References} \\
\noindent Cohen, M (2015) {\it Paradigm Shifts} Exeter: Imprint Academic. \\
\noindent Gardner, M (2001) {\it A Skeptical Look at Karl Popper} Skeptical Inquirer, 25(4):13-14,
72. \\
\noindent Kuhn, T.S. (1962, 2nd edn 1970) {\it The Structure of Scientific Revoultions} Chicago:
University of Chicago Press. \\
\noindent Ladyman, J (2002) {\it Understanding Philosohy of Science} London: Routledge.  \\
\noindent Lelli, F., McGaugh, S.~S., \& Schombert, J.~M.\ 2016, AJ, 152, 157. \\
\noindent McGaugh, S.~S., Schombert, J.~M., Bothun, G.~D., et al.\
2000, ApJ Letters, 533, L99 \\
\noindent McGaugh, S.~S., Lelli, F., \& Schombert, J.~M.\ 2016, Physics
Review Letters, 117, 201101 \\
\noindent Milgrom, M.\ 1983, ApJ, 270, 365 \\
\noindent Oemler, A.\ 1988, The Minnesota Lectures on Clusters of Galaxies, 19. \\
\noindent Popper, K. (1959) {\it Conjectures and Refutations}, London: Routledge and Kegan Paul.
\\
\noindent Sanders, R.~H.\ 2010, {\it The Dark Matter Problem: A Historical
Perspective} Cambridge University Press. \\

\end{document}